\documentclass[aps,prd,twocolumn,showpacs,nofootinbib]{revtex4-1}

\usepackage{graphicx}
\usepackage{amsmath}
\usepackage{amssymb}
\usepackage{bm}
\usepackage{bbm}
\usepackage{color}
\usepackage{slashed}
\usepackage[colorlinks=true,linkcolor=blue,citecolor=blue, urlcolor=blue]{hyperref}

\begin{document}

\title{Fragmentation functions for a quark into a spin-singlet quarkonium: Different flavor case}

\author{Xu-Chang Zheng}
\email{zhengxc@cqu.edu.cn}
\author{Ze-Yang Zhang}
\email{zhangzeyang@cqu.edu.cn}
\author{Xing-Gang Wu}
\email{wuxg@cqu.edu.cn}

\affiliation{Department of Physics, Chongqing University, Chongqing 401331, P.R. China.}

\begin{abstract}

In the paper, we calculate the fragmentation functions for a quark to fragment into a spin-singlet quarkonium, where the flavor of the initial quark is different from that of the constituent quark in the quarkonium. The ultraviolet divergences in the phase space integral are removed through the operator renormalization under the modified minimal subtraction scheme. The fragmentation function $D_{q \to \eta_Q}(z,\mu_F)$ is expressed as a two-dimensional integral. Numerical results for the fragmentation functions of a light quark or a bottom quark to fragment into the $\eta_c$ are presented. As an application of those fragmentation functions, we study the processes $Z \to \eta_c+q\bar{q}g(q=u,d,s)$ and $Z \to \eta_c+b\bar{b}g$ under the fragmentation and the direct nonrelativistic QCD approaches.

\end{abstract}


\maketitle

\section{Introduction}

According to QCD factorization theorem, the cross section for the inclusive production of a hadron $H$ with high transverse momentum ($p_T$) in a high-energy collision is dominated by the single parton fragmentation \cite{Collins:1989gx}, i.e.,
\begin{eqnarray}
 d \sigma_{A+B \to H+X}(p_T)= && \sum_i d \hat{\sigma}_{A+B\to i+X}(p_T/z,\mu_F) \nonumber \\
&& \otimes D_{i\to H}(z,\mu_F)+{\cal O}(m_H^2/p_T^2),
\label{pqcdfact}
\end{eqnarray}
where $\otimes$ denotes a convolution in the momentum fraction $z$, the sum extends over all species of partons. $d \hat{\sigma}_{A+B \to i+X}$ indicates the partonic cross section that can be calculated in perturbation theory, while $D_{i\to H}$ indicates the fragmentation function for the parton $i$ into a hadron $H$. $\mu_F$ denotes the factorization scale which is introduced to separate the energy scales of the two parts.

The factorization formula (\ref{pqcdfact}) was first derived by Collins and Soper for light hadron production \cite{Collins:1981uw}. This factorization formula can be equally applied to the heavy quarkonium production. The proof of the factorization formula (\ref{pqcdfact}) for the quarkonium production was presented by Nayak, Qiu, and Sterman \cite{Nayak:2005rt}. The factorization formula (\ref{pqcdfact}) is called leading power (LP) factorization because it gives the LP contribution in the expansion in powers of $m_H/p_T$. The factorization formula for the next-to-leading power (NLP) correction was derived in Refs.\cite{Kang:2011zza,Kang:2011mg,Fleming:2012wy,Fleming:2013qu}, and the NLP contribution comes from the double-parton fragmentation.

Fragmentation functions play an important role in the calculation of the cross sections under the LP factorization. Unlike the fragmentation functions for the production of the light hadrons which are nonperturbative in nature, the fragmentation functions for the heavy quarkonium production can be calculated through the nonrelativistic QCD (NRQCD) factorization \cite{nrqcd}. Under NRQCD factorization, the fragmentation functions for a parton to fragment into a quarkonium can be written as
\begin{eqnarray}
D_{i\to H}(z,\mu_F)=\sum_n d_{i\to (Q\bar{Q})[n]}(z,\mu_F) \langle {\cal O}^H(n) \rangle,
\label{frag.nrqcd}
\end{eqnarray}
where $d_{i\to (Q\bar{Q})[n]}$ are short-distance coefficients (SDCs) which can be expanded as powers of $\alpha_s(m_Q)$, and $\langle {\cal O}^H(n) \rangle$ are long-distance matrix elements (LDMEs).

The fragmentation functions for quarkonia have been studied extensively. Most of the fragmentation functions for the S-wave and P-wave quarkonia are known up to $\alpha_s^2$ order \cite{Chang:1992bb, Braaten:1993jn, Braaten:1993mp, Braaten:1993rw, Braaten:1994kd, Braaten:1995cj, Chen:1993ii, Yuan:1994hn, Ma:1994zt, Ma:1995ci, Ma:1995vi, Cho:1994gb, Beneke:1995yb, Braaten:2000pc, Hao:2009fa,Sang:2009zz, Jia:2012qx, Bodwin:2014bia, Ma:2013yla, Ma:2015yka, Yang:2019gga}, and a few fragmentation functions for quarkonia were calculated up to $\alpha_s^3$ order \cite{Artoisenet:2014lpa, Sepahvand:2017gup, Artoisenet:2018dbs, Feng:2018ulg, Zhang:2018mlo, Zheng:2019dfk, Zheng:2019gnb, Feng:2017cjk}. Among these studies, the next-to-leading order (NLO) corrections to the fragmentation functions for $g \to Q\bar{Q}[^1S_0^{[1,8]}]$ have been calculated recently by three groups~\cite{Artoisenet:2014lpa, Artoisenet:2018dbs, Feng:2018ulg, Zhang:2018mlo}. The NLO fragmentation functions for $g \to Q\bar{Q}[^1S_0^{[1,8]}]$ are important in prediction of the production of the $\eta_{c,b}$ and the $h_{c,b}$ at the LHC. However, the fragmentation functions for a quark into the $\eta_{c,b}$ are only available up to $\alpha_s^2$ order. Those fragmentation functions are also important to the precision prediction of the $\eta_{c,b}$ production at the LHC. Moreover, for the production of the $\eta_{c,b}$ in $e^+e^-$ collisions, the quark fragmentation contribution is more important than the gluon fragmentation contribution due to the fact that the cross section for a quark is $\alpha_s^0$ order but the cross section for a gluon is $\alpha_s$ order. In this paper, we will calculate the fragmentation functions for a quark $q$ into the $\eta_Q$, where $Q=c,b$ but $q \neq Q$.

In the early calculations of fragmentation functions for doubly heavy mesons~\cite{Chang:1992bb, Chen:1993ii}, the fragmentation functions are determined through comparing the cross section calculated based on the NRQCD factorization with that calculated based on the factorization formula (\ref{pqcdfact}) for a process containing the doubly heavy meson being produced. In fact, the fragmentation functions can be defined through the matrix elements of nonlocal gauge-invariant operators \cite{Collins:1981uw}. The operator definition for the fragmentation functions was first applied to calculations of the fragmentation functions for doubly heavy mesons by Ma \cite{Ma:1994zt}. The calculations based on the operator definition are particularly convenient to extend to higher orders. Therefore, we will calculate the fragmentation functions based on the operator definition suggested by Collins and Soper.

The paper is organized as follows. Following the Introduction, in Sec.II, we present the definition and the analytical calculation for the fragmentation functions. In Sec.III, we present the numerical results for the fragmentation functions $D_{q \to\eta_c}(z,\mu_F)(q=u,d,s)$ and $D_{b \to \eta_c}(z,\mu_F)$, and apply the fragmentation functions to processes $Z \to \eta_c+q\bar{q}g(q=u,d,s)$ and $Z \to \eta_c+b\bar{b}g$. Section IV is reserved for a summary.

\section{The analytical calculation for the fragmentation functions}
\label{analy}

\subsection{The definition of fragmentation function}

The fragmentation functions are usually defined in the light-cone coordinate system. In this coordinate system, a d-dimensional vector $V$ is expressed as $V^{\mu}=(V^+,V^-,{\bf V}_{\perp})$, with $V^+=(V^0+V^{d-1})/\sqrt{2}$ and $V^-=(V^0-V^{d-1})/\sqrt{2}$. Then the product of two vectors is $V \cdot W= V^+ W^- + V^- W^+ - {\bf V}_{\perp} \cdot {\bf W}_{\perp}$. The gauge-invariant fragmentation function for a quark $q$ to fragment into a hadron $H$ is defined as \cite{Collins:1981uw}
\begin{widetext}
\begin{eqnarray}
D_{q\to H}(z)=&&\frac{z^{d-3}}{2\pi}\sum_{X} \int dx^- e^{-iP^+ x^-/z} \nonumber \\
&&\times \frac{1}{N_c} {\rm Tr}_{\rm color}  \frac{1}{4} {\rm Tr}_{\rm Dirac} \left\lbrace \gamma^+ \langle 0 \vert \Psi(0)\bar{{\cal P}} {\rm exp}\left[ig_s \int_{0}^{\infty} dy^- A_a^+(0^+,y^-,{\bf 0}_\perp)t_a^T \right]\vert H(P^+,{\bf 0}_\perp)+X \rangle \right. \nonumber \\
&&\left. \times\langle H(P^+,{\bf 0}_\perp)+X\vert {\cal P} {\rm exp}\left[-ig_s \int_{x^-}^{\infty} dy^- A_a^+(0^+,y^-,{\bf 0}_\perp)t_a^T \right] \bar{\Psi}(x)\vert 0\rangle\right\rbrace,
\label{defrag1}
\end{eqnarray}
\end{widetext}
where $\Psi$ is the field of initial quark, $A_a^{\mu}$ is the gluon field, and $t_a(a=1 \cdots 8)$ are $SU(3)$-color matrices. The longitudinal momentum fraction is defined as $z \equiv P^+/K^+$, where $K$ is the momentum of the initial quark. The fragmentation function is defined in a reference frame in which the transverse momentum of the hadron $H$ vanishes. It is convenient to introduce a lightlike momentum whose expression is $n^{\mu}=(0,1,{\bf 0}_{\perp})$ in the reference frame where the definition of the fragmentation function carried out. Then, $z$ can be expressed as a Lorentz invariant, i.e., $z=P\cdot n/K \cdot n$. The Feynman rules can be derived from the definition (\ref{defrag1}) directly, and we have presented the Feynman rules in a previous paper \cite{Zheng:2019gnb}.

\subsection{The calculation of fragmentation function}

The definition (\ref{defrag1}) is gauge invariant. However, for the practical calculation, the gauge should be specified. We adopt the usual Feynman gauge throughout the paper. There are ultraviolet (UV) divergences in the calculation. To deal with the UV divergences, we adopt dimensional regularization with $d=4-2\epsilon$, then the UV divergences appear as the pole terms in $\epsilon$.

In the calculation, we first calculate the fragmentation function for an on-shell $Q\bar{Q}$ pair in $^1S_0^{[1]}$ state. Then the fragmentation function $D_{q \to \eta_Q}$ can be obtained from $D_{q \to (Q\bar{Q})[^1S_0^{[1]}]}$ through replacing the LDME $\langle {\cal O}^{(Q\bar{Q})[^1S_0^{[1]}]}(^1S_0^{[1]}) \rangle$ by $\langle {\cal O}^{\eta_Q}(^1S_0^{[1]}) \rangle$.

\begin{figure}[htbp]
\includegraphics[width=0.45\textwidth]{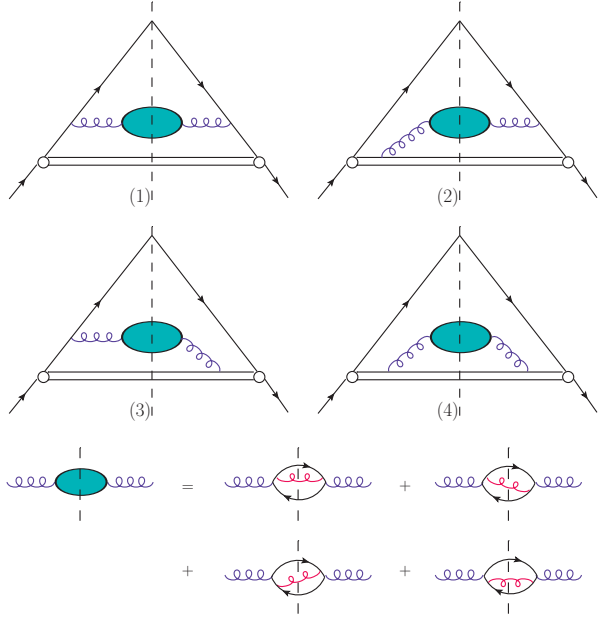}
\caption{The cut diagrams for the fragmentation function $D_{q \to (Q\bar{Q})[^1S_0^{[1]}]}$, where $q \neq Q$.
 } \label{feyn}
\end{figure}

There are 16 cut diagrams for $q(K) \to (Q\bar{Q})[^1S_0^{[1]}](p_1)+g(p_2)+q(p_3)$ under the Feynman gauge, which can be collectively represented by four diagrams in Fig.\ref{feyn}. The squared amplitudes, corresponding to four diagrams in Fig.\ref{feyn}, can be written as
\begin{eqnarray}
{\cal A}_1=&& {\rm tr}\left[\slashed{n}\frac{-i}{\slashed{p_1}+\slashed{p_2}+\slashed{p_3}-m_q-i\epsilon}(ig_s\gamma_{\nu}t^b)(\slashed{p}_{3}+m_q)\right. \nonumber \\
&& \left. (-ig_s\gamma_{\mu}t^a)\frac{i}{\slashed{p_1}+\slashed{p_2}+\slashed{p_3}-m_q+i\epsilon}\right]X_{ab}^{\mu \nu},\\
{\cal A}_2=&& {\rm tr}\left[\slashed{n}\frac{-i}{\slashed{p_1}+\slashed{p_2}+\slashed{p_3}-m_q-i\epsilon}(ig_s\gamma_{\nu}t^b)(\slashed{p}_{3}+m_q)\right. \nonumber \\
&& \left. \frac{i}{(p_1+p_2)\cdot n+i\epsilon}(ig_s n_{\mu}t^a)\right]X_{ab}^{\mu \nu},\\
{\cal A}_3=&& {\rm tr}\left[\slashed{n}(-ig_s n_{\nu}t^b)\frac{-i}{(p_1+p_2)\cdot n-i\epsilon}(\slashed{p}_{3}+m_q)\right. \nonumber \\
&& \left. (-ig_s\gamma_{\mu}t^a)\frac{i}{\slashed{p_1}+\slashed{p_2}+\slashed{p_3}-m_q+i\epsilon}\right]X_{ab}^{\mu \nu},
\end{eqnarray}
\begin{eqnarray}
{\cal A}_4=&& {\rm tr}\left[\slashed{n}(-ig_s n_{\nu}t^b)\frac{-i}{(p_1+p_2)\cdot n-i\epsilon}(\slashed{p}_{3}+m_q)\right. \nonumber \\
&& \left. \frac{i}{(p_1+p_2)\cdot n+i\epsilon}(ig_s n_{\mu}t^a)\right]X_{ab}^{\mu \nu},
\end{eqnarray}
where $\Pi_1$ is the spin-singlet projector
\begin{eqnarray}
\Pi_1=\frac{1}{(2\,m_Q)^{3/2}}(\slashed{p}_1/2-m_Q)\gamma_5(\slashed{p}_1/2+m_Q),
\end{eqnarray}
$\Lambda_1$ is the color-singlet projector
\begin{eqnarray}
\Lambda_1=\frac{\textbf{1}}{\sqrt{3}},
\end{eqnarray}
where $\textbf{1}$ is the unit matrix of the $SU(3)_c$ group. There is a common factor $X_{ab}^{\mu \nu}$ which arises from the annihilation of a virtual gluon into a $(Q\bar{Q})[^1S_0^{[1]}]$ pair and a real gluon. The factor $X_{ab}^{\mu \nu}$ can be expressed as
\begin{eqnarray}
X_{ab}^{\mu \nu}=-g_{\rho \sigma}J^{\mu \rho}_{ac}J^{\nu \sigma * }_{bc},
\end{eqnarray}
where
\begin{eqnarray}
J^{\mu \rho}_{ac}=&&{\rm tr} \left \{ \Pi_1 \Lambda_1 \left[(-ig_s\gamma_{\rho}t^c)\frac{i}{\slashed{p}_{1}/2+\slashed{p}_{2}-m_Q} (-ig_s\gamma_{\mu}t^a) \right. \right. \nonumber \\
&&\left. \left. +(-ig_s\gamma_{\mu}t^a) \frac{i}{-\slashed{p}_{1}/2-\slashed{p}_{2}-m_Q}(-ig_s\gamma_{\rho}t^c) \right] \right\}\nonumber \\
&& \cdot \frac{-i}{(p_1+p_2)^2+i \epsilon}.
\end{eqnarray}

We employ the package FeynCalc \cite{Mertig:1990an,Shtabovenko:2016sxi} to carry out the Dirac and color traces, and then the total squared amplitude (${\cal A}_{(1-4)} \equiv \sum_{i=1}^{4} {\cal A}_i$) can be written as
\begin{eqnarray}
{\cal A}_{(1-4)}=&& \frac{c_1(s_1,y,z)\, (p_3 \cdot \tilde{p})^2}{s_1^2 (s-m_q^2)^2}+\frac{c_2(s_1,y,z)\, p_1 \cdot p_3}{s_1 (s-m_q^2)^2}\nonumber \\
&& +\frac{c_3(s_1,y,z)\, p_2 \cdot p_3}{s_1 (s-m_q^2)^2}+\frac{c_4(s_1,y,z)}{(s-m_q^2)^2},
\end{eqnarray}
where
\begin{eqnarray}
&& s_1=(p_1+p_2)^2, ~~s=(p_1+p_2+p_3)^2,\nonumber \\
&& y=\frac{(p_1+p_2)\cdot n}{K\cdot n}, ~ \tilde{p}=p_1-\frac{z \,p_2}{(y-z)}.
\end{eqnarray}
The coefficients $c_i(s_1,y,z)$ can be easily extracted, and we do not list their expressions here.

The differential phase space for the fragmentation function $D_{q \to (Q\bar{Q})[^1S_0^{[1]}]}$ is\footnote{Here, we associate the scale factor $\mu^{4-d}$ with each dimensionally regulated integration in $d$ space-time dimensions. In our previous papers \cite{Zheng:2019dfk,Zheng:2019gnb}, this scale factor was put in the squared amplitudes.}
\begin{eqnarray}
d\phi_{3}(p_1,p_2,p_3)=&&2\pi \delta\left(K^+ - \sum_{i=1}^{3}  p_i^+\right)\mu^{2(4-d)}  \nonumber \\
&&\times \prod_{i=2,3}\frac{\theta(p_i^+)dp_i^+}{4 \pi p_i^+}\frac{ d^{d-2}\textbf{p}_{i\perp}}{(2\pi)^{d-2}}.
\end{eqnarray}
The contributions from the cut diagrams shown in Fig.\ref{feyn} can be calculated through
\begin{eqnarray}
D^{(1-4)}_{q \to (Q\bar{Q})[^1S_0^{[1]}]}(z)=N_{CS}\int d\phi_{3}(p_1,p_2,p_3) {\cal A}_{(1-4)},
\label{cut-contribution}
\end{eqnarray}
where $N_{CS}\equiv z^{d-3}/(8\pi N_c)$ is a factor from the definition of the fragmentation function. The integral on the right-hand side of Eq.(\ref{cut-contribution}) is UV divergent with $d=4$. This UV divergence is regularized by dimensional regularization with $d=4-2\epsilon$, and the integral generates $1/\epsilon$ terms. To perform the integration in Eq.(\ref{cut-contribution}), it is important to choose proper parametrization for the phase space. We present a parametrization for the phase space in Appendix \ref{Ap.phs}.

The differential phase space given in Eq.(\ref{eqa8}) can be expressed as follows
\begin{eqnarray}
&&N_{CS} d\phi_{3}(p_1,p_2,p_3)\nonumber \\
&& = N_{g}(p_1,p_2) d\phi_{2}(p_1,p_2) d\phi^{(3)}(p_1,p_2,p_3),
\label{eq.fact}
\end{eqnarray}
where $N_{g}(p_1,p_2)$ is defined as
\begin{eqnarray}
N_{g}(p_1,p_2)=\frac{(z/y)^{1-2\epsilon}}{(N_c^2-1)(2-2\epsilon)(2\pi\, y\, K\cdot n)},
\label{eq.Ng}
\end{eqnarray}
and $d\phi_{2}(p_1,p_2)$ is defined as
\begin{eqnarray}
d\phi_{2}(p_1,p_2)=&&\frac{z^{-1+\epsilon}(y-z)^{-\epsilon}\mu^{2\epsilon}}{2(4\pi)^{1-\epsilon}\Gamma(1-\epsilon)K\cdot n}\nonumber \\
 &&\times \left(s_1-\frac{y}{z}4m_Q^2\right)^{-\epsilon}ds_1,\label{phsp2}
\end{eqnarray}
where the range of $s_1$ is from $(4m_Q^2 y/z)$ to $\infty$. $d\phi_{2}(p_1,p_2)$ stands for the differential phase space for a gluon with longitudinal momentum $y K\cdot n$ to fragment into a $(Q\bar{Q})[^1S_0^{[1]}]$-pair with longitudinal momentum $z K\cdot n$ at leading order (LO). According to Eqs.(\ref{eq.fact}), (\ref{eq.Ng}), (\ref{phsp2}) and (\ref{eqa8}), the expression of $d\phi^{(3)}(p_1,p_2,p_3)$ can be derived
\begin{eqnarray}
&&d\phi^{(3)}(p_1,p_2,p_3)\nonumber \\
&&=\frac{(N_c^2-1)(2-2\epsilon)\mu^{2\epsilon} K\cdot n}{16N_c(2\pi)^{3-2\epsilon}} y^{1-\epsilon}(1-y)^{-\epsilon}  \nonumber \\
&&~~~\times \left[ s-s_1/y-m_q^2/(1-y)\right]^{-\epsilon}ds\, dy\, d\Omega_{3\perp}.\label{phsp4}
\end{eqnarray}
The range of $y$ is from $z$ to 1, and the range of $s$ is from $[s_1/y+m_q^2/(1-y)]$ to $\infty$.

The integrations over $\Omega_{3\perp}$ and $s$ of ${\cal A}_{(1-4)}$ can be performed using the method introduced in Ref.\cite{Zheng:2019gnb}. Then, we obtain
\begin{eqnarray}
D^{(1-4)}_{q \to (Q\bar{Q})[^1S_0^{[1]}]}(z)=&&\frac{(4\pi \mu^2)^{\epsilon}\Gamma(\epsilon)}{(4\pi)^2}\int_z^1 dy\nonumber \\
&&\int N_g d\phi_{2}(p_1,p_2) f(s_1,y,z),
\label{cut-contribution2}
\end{eqnarray}
where $N_g d\phi_{2}(p_1,p_2)\equiv N_g(p_1,p_2) d\phi_{2}(p_1,p_2)$. The expression of $f(s_1,y,z)$ is given in Appendix \ref{Ap.f}.

The contribution $D^{(1-4)}_{q \to (Q\bar{Q})[^1S_0^{[1]}]}(z)$ contains a UV pole, it should be removed through the operator renormalization \cite{Mueller:1978xu}. We carry out the renormalization using the $\overline{\rm MS}$ procedure. Then the fragmentation function under the $\overline{\rm MS}$ scheme can be obtained through
\begin{eqnarray}
&&D_{q \to (Q\bar{Q})[^1S_0^{[1]}]}(z,\mu_F)\nonumber \\
&&=D^{(1-4)}_{q \to (Q\bar{Q})[^1S_0^{[1]}]}(z)-\frac{\alpha_s}{2\pi}\left[\frac{1}{\epsilon_{UV}}- \gamma_E+ {\rm ln}~(4\pi)+{\rm ln}\frac{\mu^2}{\mu_F^2} \right]\nonumber \\
&& ~~~ \times \int_z^1 \frac{dy}{y}P_{gq}(y)D_{g\to (Q\bar{Q})[^1S_0^{[1]}]}^{\rm LO}(z/y),
\label{Dq2QQ}
\end{eqnarray}
where $\mu_F$ is the factorization scale, the expression of the splitting function $P_{gq}(y)$ is
\begin{eqnarray}
P_{gq}(y)=C_F \frac{1+(1-y)^2}{y},
\end{eqnarray}
and $D_{g\to (Q\bar{Q})[^1S_0^{[1]}]}^{\rm LO}$ is the LO fragmentation function in $d$-dimensional space-time. In the calculation, it is convenient to use the unintegrated form of $D_{g\to (Q\bar{Q})[^1S_0^{[1]}]}^{\rm LO}$, i.e.,
\begin{eqnarray}
D_{g\to (Q\bar{Q})[^1S_0^{[1]}]}^{\rm LO}(z/y)=\int N_g d\phi_{2}(p_1,p_2){\cal A}_{g \to (Q\bar{Q})[^1S_0^{[1]}]},\nonumber \\
\label{Dg2QQ}
\end{eqnarray}
where the expression of ${\cal A}_{g \to (Q\bar{Q})[^1S_0^{[1]}]}$ is
\begin{eqnarray}
&&{\cal A}_{g \to (Q\bar{Q})[^1S_0^{[1]}]}\nonumber \\
&&= \frac{16g_s^4}{3}(1-2\epsilon)(y\, K\cdot n)^2\left[\frac{(1-z/y)}{m_Q^3 s_1}-\frac{(1-z/y)}{m_Q^3 (s_1-4m_Q^2)}\right. \nonumber \\
&& ~~~ \left. +\frac{2(1-\epsilon)}{m_Q s_1^2}+\frac{4(1-z/y)^2}{m_Q (s_1-4m_Q^2)^2}\right].
\end{eqnarray}
The integral over $s_1$ in Eq.(\ref{Dg2QQ}) can be carried out easily, and we have checked that our result for $D_{g\to (Q\bar{Q})[^1S_0^{[1]}]}^{\rm LO}$ is consistent with that obtained in Refs.\cite{Braaten:1993rw,Artoisenet:2014lpa,Feng:2018ulg,Zhang:2018mlo}.

Applying Eqs.(\ref{cut-contribution2}) and (\ref{Dg2QQ}) to Eq.(\ref{Dq2QQ}), we can obtain the fragmentation function under the $\overline{\rm MS}$ scheme. It is found that the UV pole of $D^{(1-4)}_{q \to (Q\bar{Q})[^1S_0^{[1]}]}(z)$ is exactly canceled by the UV pole of the counter term from the operator renormalization. The remaining integrals no longer generate divergence, we can set $\epsilon=0$ before carrying out the integrations. Multiplying the fragmentation function $D_{q \to (Q\bar{Q})[^1S_0^{[1]}]}(z,\mu_F)$ for the $(Q\bar{Q})[^1S_0^{[1]}]$ pair by a factor $\langle {\cal O}^{\eta_Q}(^1S_0^{[1]}) \rangle/\langle {\cal O}^{(Q\bar{Q})[^1S_0^{[1]}]}(^1S_0^{[1]}) \rangle \approx \vert R_S^{(Q\bar{Q})}(0) \vert^2/(4\pi)$, we obtain the fragmentation function $D_{q \to \eta_Q}(z,\mu_F)$, i.e.,
\begin{eqnarray}
&&D_{q \to \eta_Q}(z,\mu_F)\nonumber \\
 &&=\int_z^1 dy \int_{4m_Q^2 y/z}^{\infty} d s_1\, g(s_1,\mu_F,y,z),
 \label{frag.q2etaQ1}
\end{eqnarray}
where
\begin{widetext}
\begin{eqnarray}
g(s_1,\mu_F,y,z)=&&\frac{\alpha_s^3\,\vert R_S^{(Q\bar{Q})}(0) \vert^2}{9\,\pi^2\, y^4 m_Q s_1^2 (s_1-4m_Q^2)^2\left[(1-y)s_1+y^2m_q^2\right]}\bigg\{ (y-1)\Big[s_1^3 \big( y^4-2y^3(z+1)+2y^2(z^2+6z+1)  \nonumber \\
&&  -12y\,z(z+1)+12z^2\big)+s_1^2 \big(2 y^4 m_q^2-4y^3(2m_Q^2(z+4)+z\, m_q^2)+4 y^2(4 m_Q^2(3z+2)+z^2 m_q^2)  \nonumber \\
&&-48y\,z\,m_Q^2 \big)+16 s_1 y^2 m_Q^2 \big(y^2 m_Q^2-y(2m_Q^2+z\,m_q^2)+2 m_Q^2\big)+32 y^4 m_Q^4 m_q^2 \Big]-\left[(1-y)s_1+y^2m_q^2\right]\nonumber \\
&&\left. \times \left[z^2(s_1-4m_Q^2 y/z)^2+(y-z)^2s_1^2 \right]\left[\big((1-y)^2+1\big){\rm ln}\left( \frac{(1-y)s_1+y^2m_q^2}{\mu_F^2}\right) +y^2\right]\right\}.
\label{frag.q2etaQ2}
\end{eqnarray}
\end{widetext}
Here, $R_S^{(Q\bar{Q})}(0)$ is the radial wave function at the origin for the $(Q\bar{Q})$ bound state.

\section{Numerical results and discussion}

In this section, we will present the numerical results for the fragmentation functions and apply the fragmentation functions to the decay widths for the $\eta_c$ production through Z boson decays.

The input parameters for the numerical calculation are taken as follows
\begin{eqnarray}
&& m_c=1.5\,{\rm GeV},\,m_b=4.9\,{\rm GeV},\,m_{_Z}=91.1876\,{\rm GeV},  \nonumber \\
&& \alpha=1/128,{\rm sin}^2\theta_{_W}=0.231, \vert R_S^{c\bar{c}}(0)\vert ^2=0.810\,{\rm GeV}^3.
\end{eqnarray}
The value of $\vert R_S^{c\bar{c}}(0)\vert ^2$ is taken from the potential model calculation \cite{pot}. For the strong coupling constant, we adopt two-loop formula as used in our previous paper \cite{Zheng:2019dfk}, where $\alpha_s(2m_c)=0.259$.

\subsection{The fragmentation functions}

\begin{figure}[htbp]
\includegraphics[width=0.45\textwidth]{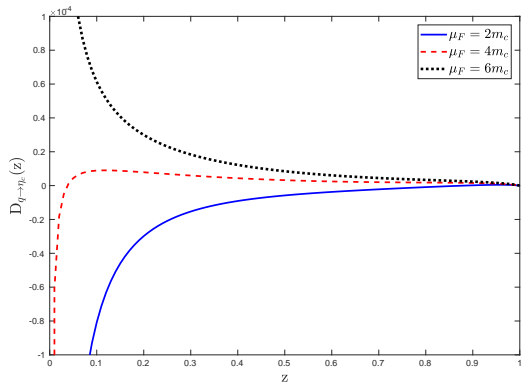}
\caption{The fragmentation function $D_{q \to \eta_c}(z,\mu_F)$ as a function of $z$ for $\mu_F=2m_c$, $4m_c$, and $6m_c$, where $q$ denotes a light quark. }
 \label{Dzqetac}
\end{figure}

The fragmentation function for a light quark into $\eta_c$, where $\mu_F=2m_c$, $4m_c$, and $6m_c$, is shown in Fig.\ref{Dzqetac}. In the numerical calculation, the mass of the light quark is neglected, and the strong coupling is taken as $\alpha_s(2m_c)$.

From Fig.\ref{Dzqetac}, we can see that the fragmentation function is sensitive to the factorization scale. When $\mu_F=2m_c$, the fragmentation function increases first ($z<0.96$) and then decreases ($z>0.96$) with the increase of $z$, and the fragmentation function is less than 0 in the most $z$ region; when $\mu_F=4m_c$, the fragmentation function also increases first ($z<0.12$) and then decreases ($z>0.12$) with the increase of $z$, but the fragmentation function is greater than 0 in the most $z$ region; When $\mu_F=6m_c$, the fragmentation function decreases monotonically with the increase of $z$,  and the fragmentation function is greater than 0 for $z \in (0,1)$. The fragmentation function has a singularity at $z=0$.

\begin{figure}[htbp]
\includegraphics[width=0.45\textwidth]{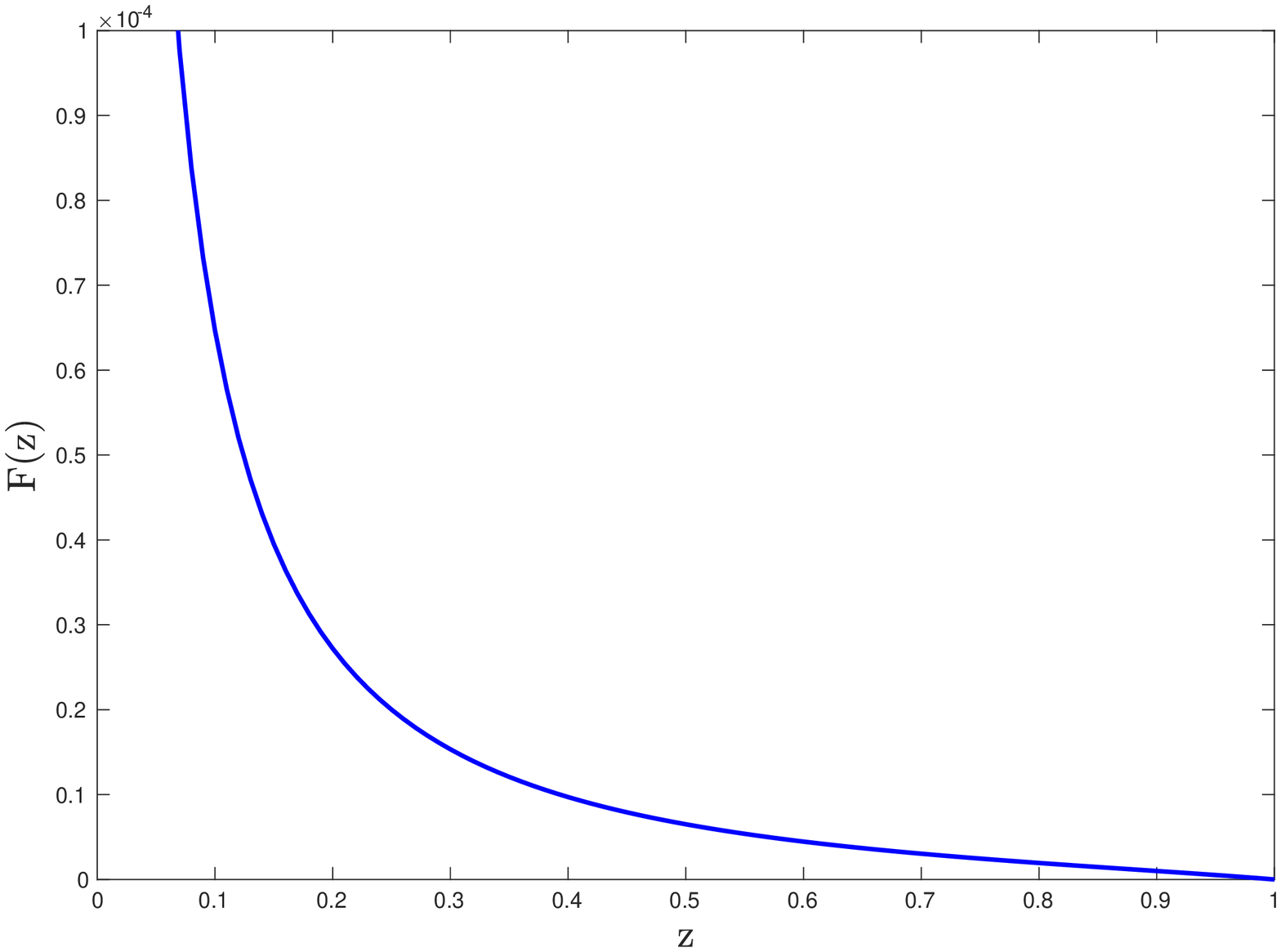}
\includegraphics[width=0.45\textwidth]{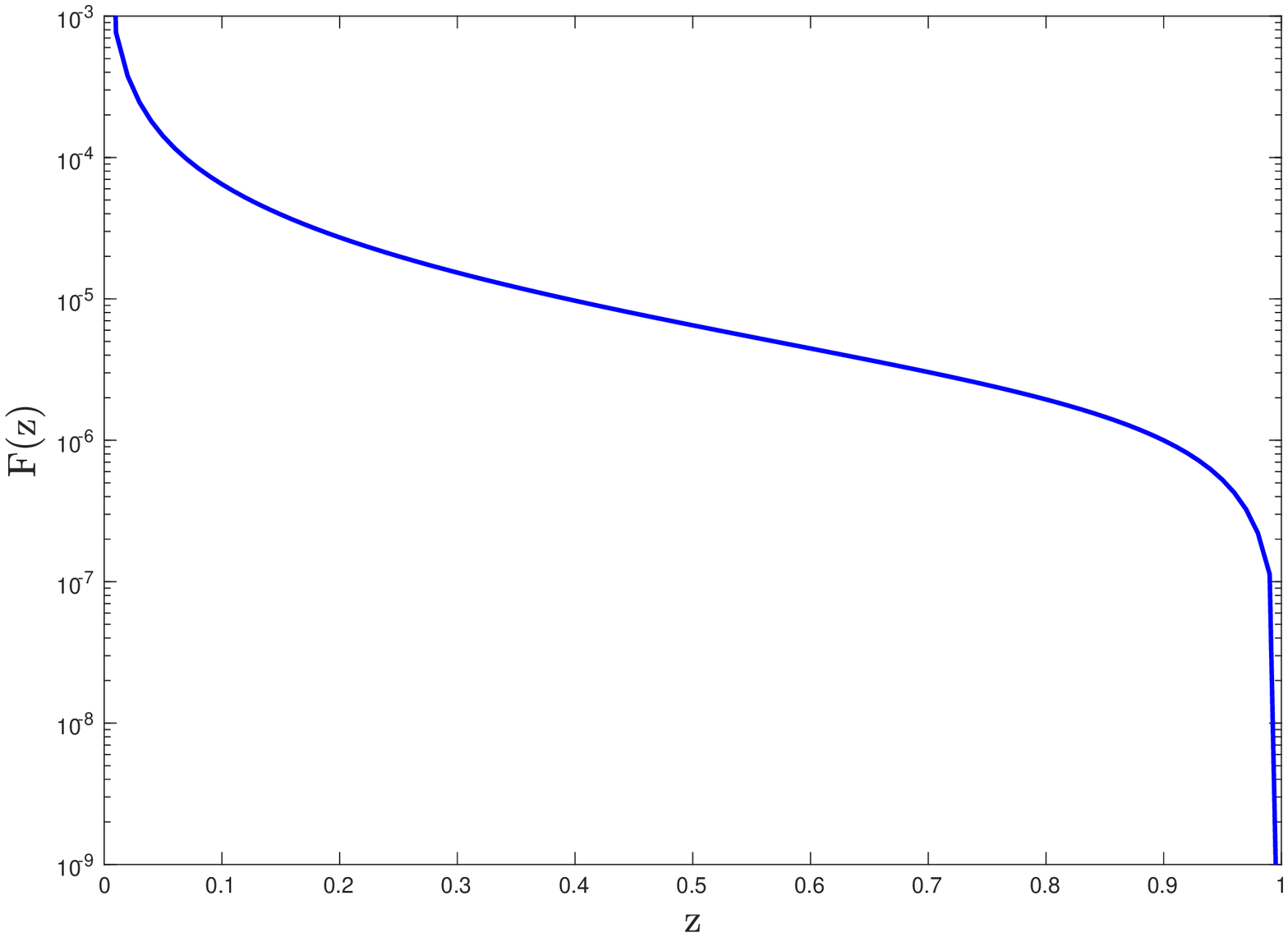}
\caption{The coefficient of ${\rm ln}(\mu_F^2/m_c^2)$ in the fragmentation function $D_{q\to \eta_c}(z,\mu_F)$, i.e., $F(z)=\frac{\alpha_s}{2\pi}\int_z^1 \frac{dy}{y}P_{gq}(y)D_{g \to \eta_c}^{LO}(z/y)$. The same curve is shown as a function of $z$ with a linear scale (upper one) and with a logarithmic scale (lower one). }
 \label{mufcoe}
\end{figure}

\begin{figure}[htbp]
\includegraphics[width=0.45\textwidth]{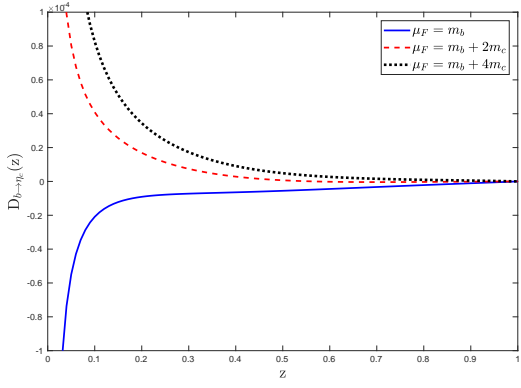}
\caption{The fragmentation function $D_{b \to \eta_c}(z,\mu_F)$ as a function of $z$ for $\mu_F=m_b$, $m_b+2m_c$ and $m_b+4m_c$.}
\label{b2etac}
\end{figure}

In order to understand the dependence of the fragmentation function on the factorization scale, we have calculated the coefficient of ${\rm ln}(\mu_F^2/m_c^2)$ in the fragmentation function $D_{q\to \eta_c}(z,\mu_F)$. The coefficient is shown as a function of $z$ in Fig.\ref{mufcoe} with a linear scale (upper one) and with a logarithmic scale (lower one). We can see that, the coefficient decreases monotonically with the increase of $z$ and is positive for $z \in (0,1)$. Like the fragmentation function, the coefficient of ${\rm ln}(\mu_F^2/m_c^2)$ also has a singularity at $z=0$. When $\mu_F$ is very large, the fragmentation function is dominated by the ${\rm ln}(\mu_F^2/m_c^2)$ term. Therefore, the behavior of the fragmentation function is similar to that of the coefficient of ${\rm ln}(\mu_F^2/m_c^2)$ when $\mu_F$ is large enough.

In Fig.\ref{b2etac}, the fragmentation function $D_{b \to \eta_c}(z,\mu_F)$ for $\mu_F=m_b$, $m_b+2m_c$ and $m_b+4m_c$ is presented. We can see that the fragmentation function $D_{b\to \eta_c}(z,\mu_F)$ is also sensitive to the factorization scale. Actually, from Eqs.(\ref{frag.q2etaQ1}) and (\ref{frag.q2etaQ2}), we can see that the coefficient of ${\rm ln}(\mu_F^2/m_Q^2)$ in $D_{b\to \eta_c}(z,\mu_F)$ is the same as that in $D_{q\to \eta_c}(z,\mu_F)$.

\subsection{Application to the decay widths}

In this subsection, we will apply the obtained fragmentation functions to the decay widths for the processes $Z \to \eta_c+q\bar{q}g$ and $Z \to \eta_c+b\bar{b}g$.

Here, we only present the calculation formulas for $Z \to \eta_c+q\bar{q}g$, the formulas for $Z \to \eta_c+b\bar{b}g$ are similar. Under the fragmentation approach, the differential decay width $d\Gamma/dz$ for $Z \to \eta_c+q\bar{q}g$ at LO can be written as
\begin{eqnarray}
\frac{d\Gamma^{\rm Frag,LO}_{Z \to \eta_c+q\bar{q}g}}{dz}=&&2\, \Gamma_{Z \to q+\bar{q}}D_{q\to \eta_c}(z,\mu_F) \nonumber \\
&& +\int^1_z \frac{dy}{y} \frac{d\hat{\Gamma}_{Z \to g+q\bar{q}}(y,\mu_F)}{dy}\nonumber \\
&& \cdot D_{g\to \eta_c}(z/y),\label{Frag.Z2etac}
\end{eqnarray}
where the energy fraction is defined as $z\equiv E_{\eta_c}/E^{\rm max}_{\eta_c}$, and $E_{\eta_c}$ and $E^{\rm max}_{\eta_c}$ are the energy and the maximum energy of the $\eta_c$ in the rest frame of the Z boson. $\Gamma_{Z \to q+\bar{q}}$ denotes the LO decay width for the process $Z \to q+\bar{q}$, and the factor of 2 is due to the fact that the contributions from the q fragmentation and $\bar{q}$ fragmentation are the same. In Eq.(\ref{Frag.Z2etac}), we have used the fact $d\hat{\Gamma}_{Z \to q+\bar{q}}/dy=\Gamma_{Z \to q+\bar{q}}\delta(1-y)$ at LO. $d\hat{\Gamma}_{Z \to g+q\bar{q}}/dy$ is the differential decay width for the inclusive production of a gluon associated with a light-quark pair. Neglecting the light quark mass, we obtain the differential decay width under the $\overline{\rm MS}$ factorization scheme
\begin{eqnarray}
&&\frac{d\hat{\Gamma}_{Z \to g+q\bar{q}}(y,\mu_F)}{dy}\nonumber \\
&&=\Gamma_{Z \to q+\bar{q}}\frac{\alpha_s}{\pi}P_{gq}(y)\left[{\rm ln}\frac{m_{_Z}^2}{\mu_F^2}+2{\rm ln}y+{\rm ln}(1-y) \right].
\end{eqnarray}
$D_{g\to \eta_c}(z)$ and $D_{q\to \eta_c}(z,\mu_F)$ are the LO fragmentation functions. The expression of $D_{g\to \eta_c}(z)$ can be found in Ref.\cite{Braaten:1993rw}, and the expression of $D_{q\to \eta_c}(z,\mu_F)$ has been given in Eq.(\ref{frag.q2etaQ1}). It is easy to check that the logarithm terms of $\mu_F^2$ in Eq.(\ref{Frag.Z2etac}) are canceled by each other which results in that $d\Gamma^{\rm Frag,LO}_{Z \to \eta_c+q\bar{q}g}/dz$ is independent of $\mu_F$.

\begin{figure}[htbp]
\includegraphics[width=0.45\textwidth]{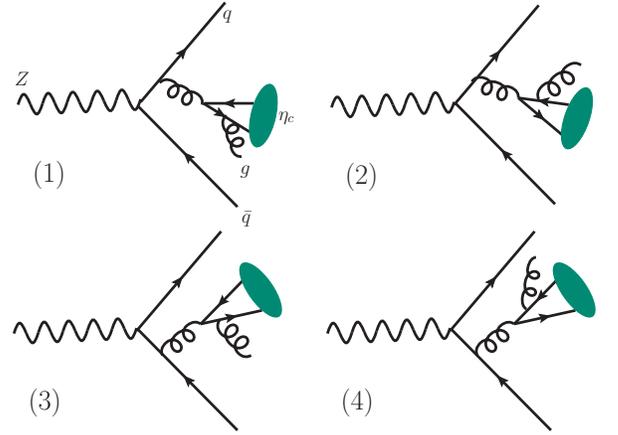}
\caption{The Feynman diagrams for $Z \to \eta_c+q\bar{q}g$ which are responsible for the fragmentation mechanism.
 } \label{feynZetacqqg}
\end{figure}

The physical picture of Eq.(\ref{Frag.Z2etac}) is as follows: The first term gives the contribution from that the Z boson decays into a light quark and a light antiquark with energies $m_{_Z}/2$ on a distance scale of order $1/m_{_Z}$, and one of the light quark and the light antiquark decays into an $\eta_c$ on a distance scale of order $1/m_c$. The second term gives the contribution from that the Z boson decays into a light quark-antiquark pair and a gluon on a distance scale of order $1/m_{_Z}$, and the gluon decays into an $\eta_c$ on a distance scale of order $1/m_c$. The two terms of Eq.(\ref{Frag.Z2etac}) share the same Feynman diagrams which are shown in Fig.\ref{feynZetacqqg}, but they come from different regions of the phase space. When the invariant mass of the virtual light quark (antiquark) is very small compared to $\mu_F$, the contribution is given by the first term of Eq.(\ref{Frag.Z2etac}). When the invariant mass of the virtual light quark (antiquark) is very large compared to $\mu_F$, the contribution is given by the second term of Eq.(\ref{Frag.Z2etac})

There are logarithms of $m_{Z}/m_c$ in the decay widths of $Z \to \eta_c+q\bar{q}g$, which may spoil the convergence of the perturbative expansion. These large logarithms can be resummed through the evolution of the fragmentation functions under the fragmentation approach. The decay width after the resummation of the leading logarithms (LLs) can be written as
\begin{eqnarray}
\frac{d\Gamma^{\rm Frag,LO+LL}_{Z \to \eta_c+q\bar{q}g}}{dz}=&&2\, \Gamma_{Z \to q+\bar{q}}D^{\rm LO+LL}_{q\to \eta_c}(z,\mu_F) \nonumber \\
&& +\int^1_z \frac{dy}{y} \frac{d\hat{\Gamma}_{Z \to g+q\bar{q}}(y,\mu_F)}{dy}\nonumber \\
&& \cdot D^{\rm LO+LL}_{g\to \eta_c}(z/y,\mu_F),\label{Frag.Z2etac.resum}
\end{eqnarray}
where the factorization scale is set as $\mu_F=m_{_Z}$, and the renormalization scale in the partonic decay widths is also set as $\mu_R=m_{_Z}$, so as to avoid large logarithms appearing in the partonic decay widths. The fragmentation functions $D^{\rm LO+LL}_{q\to \eta_c}(z,\mu_F=m_{_Z})$ and $D^{\rm LO+LL}_{g\to \eta_c}(z,\mu_F=m_{_Z})$ are obtained through solving the Dokshitzer-Gribov-Lipatov-Altarelli-Parisi (DGLAP) equations \cite{dglap1, dglap2, dglap3} with LO splitting functions, where the initial fragmentation functions $D_{q\to \eta_c}(z,\mu_{F0})$ and $D_{g\to \eta_c}(z,\mu_{F0})$ at $\mu_{F0}=2m_c$ \footnote{For $Z \to \eta_c+b\bar{b}g$ case, the initial factorization scale is taken as $\mu_{F0}=m_b+2m_c$.} are used as the boundary condition. We solve the DGLAP equations by using the program FFEVOL \cite{Hirai:2011si}.

In addition to the fragmentation approach, we can also calculate the decay width directly based on the NRQCD factorization, i.e.,
\begin{eqnarray}
d\Gamma^{\rm Direct,LO}_{Z \to \eta_c+q\bar{q}g}=d\tilde{\Gamma}_{Z \to (c\bar{c})[^1S_0^{[1]}]+q\bar{q}g}\langle {\cal O}^{\eta_c}(^1S_0^{[1]}) \rangle,
\end{eqnarray}
where we use ``Direct" to denote the results from the direct calculation based on the NRQCD factorization. The dominant contributions to the decay widths for $Z \to \eta_c+q\bar{q}g$ and $Z \to \eta_c+b\bar{b}g$ come from four fragmentation diagrams shown in Fig.\ref{feynZetacqqg}. The contributions from the nonfragmentation diagrams are suppressed by powers of $m_c/m_Z$ compared to the fragmentation contributions. For simplicity, under the  direct NRQCD calculation, we only consider the contributions from the four fragmentation diagrams shown in Fig.\ref{feynZetacqqg}.

\begin{figure}[htbp]
\includegraphics[width=0.45\textwidth]{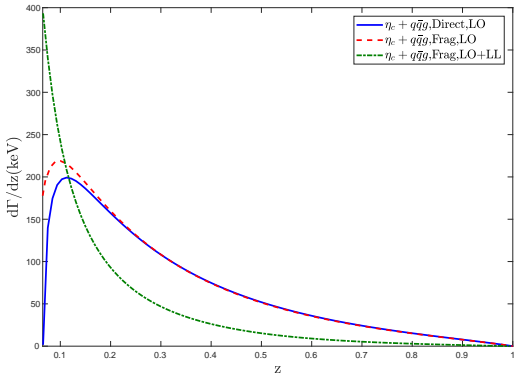}
\caption{The differential decay width $d\Gamma/dz$ as a function of $z$ for the process $Z \to \eta_c+q\bar{q}g$ under the fragmentation and the direct NRQCD calculations. The contributions for $q=u,d,s$ are summed.}
\label{gammazq}
\end{figure}

\begin{figure}[htbp]
\includegraphics[width=0.45\textwidth]{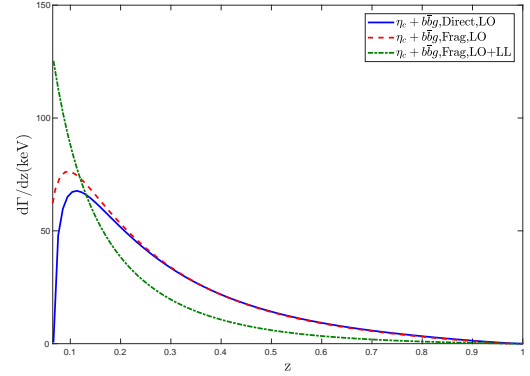}
\caption{The differential decay width $d\Gamma/dz$ as a function of $z$ for $Z \to \eta_c+b\bar{b}g$ under the fragmentation and the direct NRQCD calculations.}
\label{gammazb}
\end{figure}

\begin{figure}[htbp]
\includegraphics[width=0.45\textwidth]{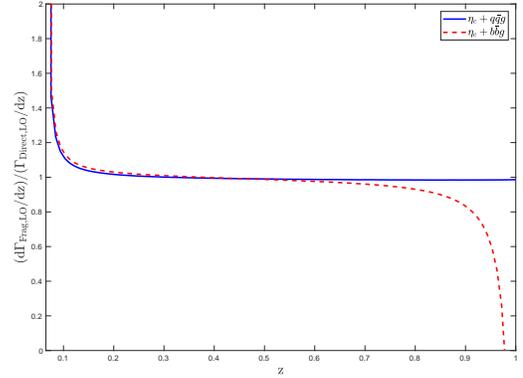}
\caption{The ratios $(d\Gamma_{\rm Frag,LO}/dz)/(d\Gamma_{\rm Direct,LO}/dz)$ as functions of $z$ for the processes $Z \to \eta_c+q\bar{q}g$ and $Z \to \eta_c+b\bar{b}g$.}
\label{gammazR}
\end{figure}

The differential decay widths for $Z \to \eta_c+q\bar{q}g$ and $Z \to \eta_c+b\bar{b}g$ under the fragmentation and the direct NRQCD approaches are presented in Figs.\ref{gammazq} and \ref{gammazb}. In order to see the difference between the ``Frag,LO" results and the ``Direct,LO" results more clearly, the ratios $(d\Gamma_{\rm Frag,LO}/dz)/(d\Gamma_{\rm Direct,LO}/dz)$ are given in Fig.\ref{gammazR}. For the ``Frag,LO" and the ``Direct,LO" calculations, the renormalization scale is fixed as $\mu_R=2m_c$ for simplicity. For the ``Frag,LO+LL" calculation, the choice of the renormalization scale and the factorization scale has been described below Eq.(\ref{Frag.Z2etac.resum}). From the figures, we can see that the differential decay widths from the ``Frag,LO" calculation are very close to those from the ``Direct,LO" calculation, especially for $0.2 \leq z \leq 0.8$.

The total decay widths can be obtained through integrating the differential decay widths $d\Gamma/dz$ over $z$. The total decay widths under the fragmentation approach and the direct NRQCD approach are given in Table \ref{tb.width}. We can see that the total decay widths obtained from the ``Frag,LO" calculation and the ``Direct,LO" calculation are also very close. Therefore, the fixed-order fragmentation approach (i.e., the ``Frag,LO" approach) provides a good approximation to the direct NRQCD calculation.

The differential and total decay widths after the resummation of the large logarithms under the fragmentation approach are also shown in Figs.\ref{gammazq} and \ref{gammazb} and Table \ref{tb.width}. We can see that, after the resummation, the differential decay widths are enhanced at smaller $z$ values but are reduced at larger $z$ values, and the total decay widths are reduced after the resummation. In fact, the fixed-order results have a big uncertainty caused by the choice of the renormalization scale. If we set the renormalization scale as $\mu_R=m_{_Z}$, the fixed-order results become $[\alpha_s(m_{_Z})/\alpha_s(2m_c)]^3=0.0958$ of those with $\mu_R=2m_c$. On the contrary, the results after the resummation have a smaller uncertainty caused by the choice of the renormalization scale. Because the renormalization scale of the initial fragmentation functions should be ${\cal O}(m_c)$, and the renormalization scale of the coefficient functions should be ${\cal O}(m_{_Z})$. Moreover, the resummed results include the leading logarithms up to all orders from the collinear radiation. Therefore, the results after the resummation are more precise than the fixed-order results.

\begin{table}[h]
\begin{tabular}{c c c}
\hline
~& $\Gamma_{Z \to \eta_c+q\bar{q}g}$ (keV) & $\Gamma_{Z \to \eta_c+b\bar{b}g}$(keV)  \\
\hline
Direct   & 63.0 &  19.3\\
Frag,LO  & 65.6 &  20.3 \\
Frag,LO+LL  & 40.3 & 15.3  \\
\hline
\end{tabular}
\caption{The decay widths of $Z \to \eta_c+q\bar{q}g$ and $Z \to \eta_c+b\bar{b}g$ under the fragmentation and the direct NRQCD approaches. For the $Z \to \eta_c+q\bar{q}g$ case, the contributions for $q=u,d,s$ are summed.}
\label{tb.width}
\end{table}

\section{Summary}

In the present paper, we have calculated the fragmentation functions for a (heavy or light) quark into a spin-singlet quarkonium, where the flavor of the initial quark is different from that of the constituent quark in the quarkonium. There are UV divergences in the phase-space integral, which are removed through the operator renormalization of the fragmentation function. We have carried out the renormalization under the $\overline{\rm MS}$ scheme. The fragmentation function $D_{q \to \eta_Q}(z,\mu_F)$ is given as a two-dimensional integral, and this two-dimensional integral can be calculated easily through numerical integration. Numerical results for a light quark or a bottom quark into the $\eta_c$ with several factorization scales are analyzed. The results show that these fragmentation functions are sensitive to the factorization scale. Especially, when $\mu_F$ is small, the fragmentation functions are negative at small $z$ values. There is a singularity at $z=0$ for these fragmentation functions.

We have applied the obtained fragmentation functions to the decay widths for the processes $Z \to \eta_c+q\bar{q}g(q=u,d,s)$ and $Z \to \eta_c+b\bar{b}g$. The differential decay widths and total decay widths are calculated under the fragmentation and the direct NRQCD approaches. It is found that the results under the fixed-order fragmentation and the direct NRQCD approaches are close to each other. Therefore, the fixed-order fragmentation approach provides a good approximation to the direct NRQCD calculation. The more precise results containing the resummation of the large logarithms under the fragmentation approach are also presented. Moreover, the fragmentation functions obtained in this paper can be used in the studies on the production of $\eta_c$ and $\eta_b$ at high-energy colliders.

\hspace{2cm}

\noindent {\bf Acknowledgments:} This work was supported in part by the Natural Science Foundation of China under Grants No. 11625520, No. 12005028 and No. 12047564, by the Fundamental Research Funds for the Central Universities under Grant No. 2020CQJQY-Z003, and by the Chongqing Graduate Research and Innovation Foundation under Grant No. ydstd1912.

\appendix

\section{The parametrization for the phase space}
\label{Ap.phs}

In order to extract the UV poles in the calculation analytically, the phase space for the fragmentation function should be parametrized properly. In this Appendix, we will present a parametrization for the phase space. The differential phase space for the fragmentation function $D_{q \to (Q\bar{Q})[^1S_0^{[1]}]}$ is
\begin{eqnarray}
d\phi_{3}(p_1,p_2,p_3)=&&2\pi \delta\left(K^+ - \sum_{i=1}^{3}  p_i^+\right)\mu^{2(4-d)}  \nonumber \\
&&\times \prod_{i=2,3}\frac{\theta(p_i^+)dp_i^+}{4 \pi p_i^+}\frac{ d^{d-2}\textbf{p}_{i\perp}}{(2\pi)^{d-2}}.\label{eqa1}
\end{eqnarray}
According to Ref.\cite{Zheng:2019gnb}, the differential phase space for a single parton with momentum $p_i$ and mass $m_i$ can be expressed as
\begin{eqnarray}
\frac{d^{d-1}\textbf{p}_i}{(2\pi)^{d-1}2p_i^0}=\frac{(\lambda_i p_i\cdot n-m_i^2)^{-\epsilon}}{4(2\pi)^{3-2\epsilon}}  d\lambda_i d (p_i\cdot n) \,d\Omega_{i\perp},\label{eqa2}
\end{eqnarray}
where
\begin{eqnarray}
\lambda_i=2 k_i\cdot p_i/k_i\cdot n,\label{eqa3}
\end{eqnarray}
and $k_i$ is an arbitrary lightlike momentum which is not parallel to $n$. $d\Omega_{i\perp}$ stands for the differential transverse solid angle, and the total transverse solid angle $\Omega_{i\perp}=2\pi^{1-\epsilon}/\Gamma(1-\epsilon)$.

Applying the parametrization (\ref{eqa2}) to the differential phase spaces for $p_2$ and $p_3$, we obtain
\begin{eqnarray}
d\phi_{3}(p_1,p_2,p_3)=&&\frac{2^{-2\epsilon}(K\cdot n)^{1-2\epsilon}\mu^{4\epsilon}}{(4\pi)^{4-3\epsilon}\Gamma(1-\epsilon)}[(1-y)(y-z)]^{-\epsilon} \nonumber \\
&&\times \left[1-\frac{m_q^2}{\lambda_3(1-y)K\cdot n}\right]^{-\epsilon}\lambda_2^{-\epsilon}\lambda_3^{-\epsilon}  \nonumber \\
&& \times dy\, d\lambda_2 \,d\lambda_3 \, d\Omega_{3\perp}, \label{eqa4}
\end{eqnarray}
where the integrations over $p_2 \cdot n$ and $\Omega_{2\perp}$ have been performed.

To obtain the phase-space parametrization used to extract the UV poles, we choose the light-like momenta $k_2$ and $k_3$ as follows:
\begin{eqnarray}
k_2^{\mu}&=&p_1^{\mu}-\frac{2m_Q^2}{p_1\cdot n}n^{\mu},  \nonumber \\
k_3^{\mu}&=&(p_1+p_2)^{\mu}-\frac{s_1}{2(p_1+p_2)\cdot n}n^{\mu},
\label{eqa5}
\end{eqnarray}
then we obtain
\begin{eqnarray}
\lambda_2=\frac{1}{zK\cdot n}\left(s_1-\frac{y}{z}4m_Q^2\right),\label{eqa6}
\end{eqnarray}
and
\begin{eqnarray}
\lambda_3=\frac{1}{y K\cdot n}\left(s-\frac{s_1}{y}-m_q^2\right),\label{eqa7}
\end{eqnarray}
Changing variables in Eq.(\ref{eqa4}) from $\lambda_2$ and $\lambda_3$ to $s_1$ and $s$, we obtain
\begin{eqnarray}
d\phi_3(p_1,p_2,p_3)=&&\frac{2^{-2\epsilon}(z\, y)^{-1+\epsilon}\mu^{4\epsilon}}{(4\pi)^{4-3\epsilon}\Gamma(1-\epsilon)K\cdot n}(1-y)^{-\epsilon} \nonumber \\
&&\times (y-z)^{-\epsilon}[s-s_1/y-m_q^2/(1-y)]^{-\epsilon} \nonumber \\
&&\times (s_1-4m_Q^2 y/z)^{-\epsilon} dy\, ds \,ds_1 \, d\Omega_{3\perp}.\label{eqa8}
\end{eqnarray}

\begin{widetext}

\section{The expression of $f(s_1,y,z)$}
\label{Ap.f}

The expression of $f(s_1,y,z)$ which appeared in Eq.(\ref{cut-contribution2}) can be written in following form:
\begin{eqnarray}
f(s_1,y,z)=\frac{2^8 g_s^6[(1-y)s_1+y^2 m_q^2]^{-\epsilon} (K\cdot n)^2}{9  m_Q s_1^2  (s_1-4m_Q^2)^2}(f_0 +\epsilon f_1),
\end{eqnarray}
where
\begin{eqnarray}
f_0=[(1-y)^2+1][s_1^2(y^2-2y\,z+2z^2)-8m_Q^2 s_1 y\,z+16 y^2 m_Q^4],
\end{eqnarray}
and
\begin{eqnarray}
f_1=&& \frac{1}{[(1-y)s_1+y^2 m_q^2]}\Big\{s_1^2\Big[s_1(y-1)\big(5y^4-8y^3(z+1)+4y^2(z+2)(2z+1)-20y\,z(z+1)+20z^2\big)\nonumber \\
&&-2y^2 m_q^2(y^2-2y+2)(2y^2-3y\,z+3z^2)\Big]-8y m_Q^2 s_1\Big[s_1(y-1)(y^3+2y^2(2z+1)-2y(5z+1)+10z)\nonumber \\
&&-y^2 m_q^2 (y^2-2y+2)(y+3z)\Big]+16m_Q^4 y^2\Big[s_1(5y^3-13y^2+16y-8)-4m_q^2 y^2(y^2-2y+2)\Big]\Big\}.
\end{eqnarray}
\end{widetext}

\end{document}